# Structural origins of the properties of rare earth nickelate superlattices

**Jinwoo Hwang[a)], Junwoo Son, Jack Y. Zhang, Anderson Janotti, Chris G. Van De Walle, and Susanne Stemmer[b)]**

Materials Department, University of California, Santa Barbara, California 93106-5050, U.S.A.

**Abstract**

$NiO_6$ octahedral tilts in the $LaNiO_3/SrTiO_3$ superlattices are quantified using position averaged convergent beam electron diffraction in scanning transmission electron microscopy. It is shown that maintaining oxygen octahedra connectivity across the interface controls the octahedral tilts in the $LaNiO_3$ layers, their lattice parameters and their transport properties. Unlike films and layers that are connected on one side to the substrate, subsequent $LaNiO_3$ layers in the superlattice exhibit a relaxation of octahedral tilts towards bulk values. This relaxation is facilitated by correlated tilts in $SrTiO_3$ layers and is correlated with the conductivity enhancement of the $LaNiO_3$ layers in the superlattices relative to individual films.

[a] Electronic mail: jhwang@mrl.ucsb.edu

[b] Electronic mail: stemmer@mrl.ucsb.edu

Superlattices with strongly correlated "Mott" materials, such as LaNiO$_3$, have generated significant interest for designing novel ground states, such as superconductivity, not present in either of the bulk constituents [1-3]. To design the properties of such superlattices, interfacial proximity effects, such as associated with coupling of the lattice properties and structural coherency across the interface, must be understood. In particular, the physical properties of strongly correlated oxides, such as the rare earth nickelates (general formula: $R$NiO$_3$, where $R$ is a trivalent rare earth ion), are sensitive to subtle deviations from the ideal cubic perovskite structure, such as tilts or distortions of the NiO$_6$ octahedra [4-6]. For example, the temperature of the metal-to-insulator transition systematically increases across the $R$NiO$_3$ series ($R \neq$ La) with increasing deviation of the Ni-O-Ni bond angle from the ideal 180° angle in the cubic perovskite structure [7]. In thin films, octahedral tilts are modified by epitaxial coherency strains [8-17]. It is, however, less well understood how (or if) they respond by interfacial coupling to the oxygen octahedral tilts in adjacent layers or the substrate.

While bulk LaNiO$_3$ is a metal at all temperatures, coherently strained LaNiO$_3$ thin films exhibit a transition to strongly localized transport below a critical thickness, usually of a few unit cells (u.c.'s) [18-20]. Several explanations have been proposed in the literature, including Anderson localization, which occurs when the *resistance* exceeds a critical value (Mott minimum conductivity) below a certain thickness [19, 21], and quantum confinement [22, 23]. The latter has also been invoked for thickness-induced transitions in other correlated materials [24]. Recent results show that ultrathin films *are metallic* when embedded in superlattice geometries [25, 26]. This suggests that subtle structural differences that are difficult to characterize by techniques that spatially average, play an essential role in determining the transport properties of ultrathin correlated films.

We have recently shown that position averaged convergent beam electron diffraction (PACBED) in scanning transmission electron microscopy (STEM) [27] has u.c. spatial resolution and is sensitive to picometer-small structural distortions [28, 29]. This makes PACBED ideally suited for a complete, spatially resolved understanding of the structural origins of superlattice transport properties. Here, we use PACBED to quantify octahedral tilts in [LaNiO$_3$/SrTiO$_3$]$_n$ superlattices, which show an insulator-to-metal transition at $n \geq 2$ [25]. We show that the enhanced electrical conductivity in superlattice geometries is due to relaxations of octahedral tilts towards bulk values. The studies reveal the relative importance of lattice mismatch and interfacial connectivity in the observed tilt patterns.

Superlattices of [4 u.c. LaNiO$_3$/3 u.c. SrTiO$_3$]$_n$ were grown on (001) (LaAlO$_3$)$_{0.3}$(SrAl$_{0.5}$Ta$_{0.5}$O$_3$)$_{0.7}$ (LSAT), as described in ref. [25]. TEM samples were prepared by 2° wedge polishing. To remove surface layers and damage, samples were wet-etched in hydrofluoric acid for 5 s. A FEI Titan S/TEM operated at 300 kV with a Gatan Enfina CCD was used for PACBED and high-angle annular dark-field (HAADF) imaging. PACBED patterns were simulated using a frozen phonon multislice algorithm [30]. All superlattices were coherently strained [25], i.e., the in-plane lattice parameter was that of the LSAT substrate (3.87 Å). The out-of-plane lattice parameter was calculated using the aspect ratio of the pseudocubic u.c. measured by PACBED. Bulk LaNiO$_3$ is rhombohedral (space group $R\bar{3}c$) with $a^-a^-a^-$ octahedral tilts in Glazer notation [4] (Fig. 1). The negative signs imply that neighboring octahedra tilt in opposite directions along each axis. The tilt angles about the $x, y$ and $z$ axes are identical ($\alpha = \beta = \gamma = 5.2°$) [31] and result in an Ni-O-Ni bond angle of 165.2°. In superlattices and thin films strained to a cubic substrate, the $x$ and $y$ directions are equivalent, thus $\alpha = \beta$, and the relative tilt direction between neighboring octahedra is preserved; thus octahedral tilt patterns in films and superlattices are of type $a^-a^-c^-$ [8, 13, 29]. For quantitative comparisons of experimental and simulated PACBED, $\chi^2$ maps were calculated as a function of $\alpha$ and $\gamma$ for possible TEM sample thicknesses, $t$ [29]. The brightest pixels have the lowest $\chi^2$ (see scale bars). The $\alpha$ and $\gamma$ values providing the best match are defined as the minimum of a 2-dimensional polynomial fit to the $\chi^2$ map. The errors stated below are those of the fit [29]. Experimental tilt patterns were compared with density functional theory (DFT) within the local spin-density approximation (LSDA) [32, 33]. DFT-LSDA underestimates the lattice parameters. The calculated lattice parameter of bulk LaNiO$_3$ was 2% smaller than the experimental value [31]. To simulate LaNiO$_3$ coherently strained to LSAT, the in-plane lattice constant ($a$) was increased by 0.78% and the out-of-plane lattice constant $c$ changed according the experimentally observed $c/a$ ratio.

Figure 2 shows experimental HAADF images, experimental and simulated (best match) PACBED patterns and the $\chi^2$ maps for each layer of a [LaNiO$_3$/SrTiO$_3$]$_2$ superlattice. From the $\chi^2$ maps, the octahedral tilt angles in the bottom LaNiO$_3$ layer are $\alpha = 2.8 \pm 0.4°$ and $\gamma = 7.1 \pm 0.6°$. The out-of-plane tilt ($\alpha$) was thus smaller, and the in-plane tilt ($\gamma$) larger, relative to bulk LaNiO$_3$. The change in tilt angles *is opposite* to what is observed for a thin film of LaNiO$_3$ on LSAT, for which $\alpha$ (6.2 ± 0.4°) is larger and $\gamma$ (0.9 ± 0.8°) is smaller relative to bulk LaNiO$_3$ [29]. This result appears to contradict what is expected if tilt patterns are a result of epitaxial strain only, since both layers are under the same coherency strain. The difference can be understood as follows. For an individual film of LaNiO$_3$ on LSAT the octahedral tilt pattern is primarily determined by the tensile in-plane strain imposed by the substrate [8, 29]. For LaNiO$_3$ embedded in the superlattice, however, the tilt pattern must also satisfy the need to maintain oxygen octahedral connectivity at both interfaces. This results in a straightening of the octahedra along the growth direction, resulting in the observed reduction of the out-of-plane angle. Further evidence that octahedral connectivity is a driving force for the observed tilt pattern comes from the analysis of the shape of the pseudocubic u.c. of the bottom $LaNiO_3$ layer in the superlattice. Figure 2(e) shows that the diffraction disks are spaced further apart in the *in-plane direction*: the out-of-plane/in-plane ($c/a$) ratio of the pseudocubic u.c. measured from PACBED is $1.01 \pm 0.003$, which gives $c = 3.91 \pm 0.01$ Å for the bottom $LaNiO_3$ layer. This elongation seems to imply a *negative Poisson's ratio*, as the in-plane strain imposed by the substrate is *tensile*. This should cause ($c/a$) < 1, as dictated by the elastic constants. As shown in Fig. 3, ($c/a$) > 1 can, however, be explained as being a result of satisfying octahedral connectivity requirements: to connect the oxygen octahedra, the out-of plane tilt angles, $\alpha$ and $\beta$, decrease to near zero (left schematic in Fig. 3). This elongates the u.c. in all directions. To satisfy coherency with the LSAT substrate lattice parameter, the $\gamma$ angle increases to ~ 7°, which decreases the *in-plane* lattice parameters (right schematic) to match those of LSAT.

The tilt angles in subsequent $LaNiO_3$ layers in the superlattice gradually relax to those of bulk $LaNiO_3$, with the topmost $LaNiO_3$ layer showing $\alpha = 4.6 \pm 0.8°$ and $\gamma = 5.4 \pm 1.4°$. PACBED patterns from a $[LaNiO_3/SrTiO_3]_3$ superlattice show a similar relaxation [34]. While similar constraints of octahedral connectivity should apply, analysis of the octahedral tilts in the $SrTiO_3$ layers [middle rows, Figs. 1(b-d)] shows that unlike bulk $SrTiO_3$, where the tilt is zero ($SrTiO_3$ is cubic at room temperature), the $SrTiO_3$ layer has $\alpha$ ~ 1.5°. The non-zero tilts in the $SrTiO_3$ layers may allow subsequent $LaNiO_3$ layers to relax their tilt angles to (nearly) bulk values. This is in contrast to the tilt patterns found in $LaNiO_3$ layers on thick $SrTiO_3$ substrates, which are similar to those on LSAT [8]. The $c/a$ ratio for the $SrTiO_3$ layer is > 1 ($1.01 \pm 0.002$), as expected, since the lattice parameter of $SrTiO_3$ (3.905 Å) is larger than that of LSAT. The $c/a$ ratio of the top LaNiO$_3$ layer is 1.0 ± 0.002. Thus the elongation along the *c*-axis seen in the bottom LaNiO$_3$ layer is reduced along with the relaxation of the octahedral tilts to bulk values. In summary, the tilt angles in superlattices arise from a combination of epitaxial strain and oxygen octahedral connectivity. In the upper layers of LaNiO$_3$/SrTiO$_3$ superlattices, this results in angles favorable for low resistivity, as discussed below.

Table I summarizes the Ni-O-Ni bond angles and distances calculated from the measured octahedral tilt angles and lattice parameters. The *out-of-plane* Ni-O-Ni angle (162 ± 1.1°) is reduced for the individual LaNiO$_3$ film [29], while the bottom LaNiO$_3$ layer in a [LaNiO$_3$/SrTiO$_3$]$_2$ superlattice has an increased angle (172 ± 1.1°). The *in-plane* Ni-O-Ni bond angles are close to bulk values, consistent with the DFT results by May et al. [8]. The calculated Ni-O bond distances are relatively unaffected. For comparison, the angles obtained from DFT simulations are also shown in Table I, and show similar trends.

We next discuss the relationships between the observed tilt patterns and the electrical transport. Figure 4 shows the inverse of the sheet resistance ($1/R_s$) of superlattices with $n \geq 2$, as a function of $n$ (superlattices with $n = 1$ were insulating [25]). A linear relationship is observed between $n$ and $1/R_s$ for all temperatures, with an intercept at $n = 1$. For layers that are connected in parallel the sheet resistance is given by:

$$\frac{1}{R_s} = \frac{1}{R_{n \geq 2}}(n-1) + \frac{1}{R_{n=1}}, \tag{1}$$

where $R_{n \geq 2}$ is the sheet resistance of all layers except the bottom layer, and $R_{n=1}$ is the sheet resistance of the bottom LaNiO$_3$ layer. If $R_{n=1}$ is large, i.e. the bottom layer is strongly localized, then the second term on the right hand side of Eq. (1) can be neglected, and plots of $1/R_s(n)$ intercept at $n = 1$, which is what is observed in the experiments. We note that alternative models, invoking interfacial layers or a percolation threshold [25], provided less satisfactory description of the observed behavior than Eq. (1) with $1/R_{n=1} \to 0$. Comparison with Table I shows that strong localization in the bottom layer in the superlattice is correlated with an *increased* out-of-plane Ni-O-Ni angle (172 ± 1.1°). All other layers have Ni-O-Ni angles close to bulk (167 ± 2.3°) and are metallic. For comparison, a single-layered LaNiO$_3$ has a *reduced* Ni-O-Ni angle (162 ± 1.1°) and is also insulating. Thus, any deviation from bulk angles appears to cause strong localization below a critical film thickness. We note that the Ni-O-Ni bond angles tabulated in Table I are not confined purely to the planes parallel and perpendicular to the substrate surface, respectively. For instance, the in-plane bond angle has an out-of-plane component. Previous studies have shown that *d*-band width and mass enhancement in LaNiO$_3$ are all correlated with film strain [19, 35, 36] – the present results show that these properties are largely dominated by the bond angles. For bulk $R$NiO$_3$ smaller Ni-O-Ni angles (as determined by the size of *R*) are associated with reduced band width, an increased metal-to-insulator transition temperature [7] and higher resistivity [37]. It may therefore appear counterintuitive that the bottom layer in the superlattice with its *increased* out-of-plane Ni-O-Ni angle should have high resistance. While it is possible that the increased bond lengths in this layer (see Table I) may play a role, future studies should clarify to what degree insights from bulk $R$NiO$_3$ apply to thin LaNiO$_3$ films for which the in- and out-plane angles differ, and the *R* cation is not varied. We also note that the strongly localized behavior is consistent with a disorder-induced Anderson transition [19], rather than the Mott metal-insulator transition of the bulk nickelates [38, 39]. Only very recently models have become available for $R$NiO$_3$ that take strong electron correlations into account [38, 39]. Correlation physics may drive new ground states in the presence of disorder and associated suppression of the kinetic energy [22, 38]. The extreme sensitivity of properties to both epitaxial strain and octahedral connectivity should allow for fine-tuning properties not possible with bulk materials.

We thank Varistha Chobpattana and Santosh Raghavan for help with the TEM sample preparation, and Jim Allen for discussions. J. H. and S.S. acknowledge support by DOE (DEFG02-02ER45994) and by the NSF-supported UCSB MRSEC (DMR-1121053), which also supported the facilities used in this study. J. S. was supported by a MURI from the Army Research Office (W911-NF-09-1-0398). A.J. was supported by ARO (W911-NF-11-1-0232). The authors also acknowledge use of computing facilities of the Center for Scientific Computing at the California Nanosystems Institute and Materials Research Laboratory in UC Santa Barbara (NSF CNS-0960316).

**References**

[1]   P. Hansmann, X. P. Yang, A. Toschi, G. Khaliullin, O. K. Andersen, and K. Held, Phys. Rev. Lett. **103**, 016401 (2009).

[2]   J. Chaloupka, and G. Khaliullin, Phys. Rev. Lett. **100**, 016404 (2008).

[3]   M. J. Han, X. Wang, C. A. Marianetti, and A. J. Millis, Phys. Rev. Lett. **107**, 206804 (2011).

[4]   A. M. Glazer, Acta Cryst. B **28**, 3384 (1972).

[5]   P. M. Woodward, Acta Crystallogr. B **53**, 32 (1997).

[6]   M. Imada, A. Fujimori, and Y. Tokura, Rev. Mod. Phys. **70**, 1039 (1998).

[7]   J. B. Torrance, P. Lacorre, A. I. Nazzal, E. J. Ansaldo, and C. Niedermayer, Phys. Rev. B **45**, 8209 (1992).

[8]   S. J. May, J. W. Kim, J. M. Rondinelli, E. Karapetrova, N. A. Spaldin, A. Bhattacharya, and P. J. Ryan, Phys. Rev. B **82**, 014110 (2010).

[9]   J. M. Rondinelli, S. J. May, and J. W. Freeland, MRS Bull. **37**, 261 (2012).

[10]  F. Z. He, B. O. Wells, Z. G. Ban, S. P. Alpay, S. Grenier, S. M. Shapiro, W. D. Si, A. Clark, and X. X. Xi, Phys. Rev. B **70**, 235405 (2004).

[11]  A. Vailionis, H. Boschker, W. Siemons, E. P. Houwman, D. H. A. Blank, G. Rijnders, and G. Koster, Phys. Rev. B **83**, 064101 (2011).

[12]  J. C. Woicik, C. K. Xie, and B. O. Wells, J. Appl. Phys. **109**, 083519 (2011).

[13]  S. J. May, C. R. Smith, J. W. Kim, E. Karapetrova, A. Bhattacharya, and P. J. Ryan, Phys. Rev. B **83**, 153411 (2011).

[14]  H. Rotella, U. Luders, P. E. Janolin, V. H. Dao, D. Chateigner, R. Feyerherm, E. Dudzik, and W. Prellier, Phys. Rev. B **85**, 184101 (2012).

[15] C. L. Jia, S. B. Mi, M. Faley, U. Poppe, J. Schubert, and K. Urban, Phys. Rev. B **79**, 081405 (2009).

[16] J. He, A. Borisevich, S. V. Kalinin, S. J. Pennycook, and S. T. Pantelides, Phys. Rev. Lett. **105**, 227203 (2010).

[17] J. Chakhalian *et al.*, Phys. Rev. Lett. **107**, 116805 (2011).

[18] R. Scherwitzl, P. Zubko, C. Lichtensteiger, and J. M. Triscone, Appl. Phys. Lett. **95**, 222114 (2009).

[19] J. Son, P. Moetakef, J. M. LeBeau, D. Ouellette, L. Balents, S. J. Allen, and S. Stemmer, Appl. Phys. Lett. **96**, 062114 (2010).

[20] S. J. May, T. S. Santos, and A. Bhattacharya, Phys. Rev. B **79**, 115127 (2009).

[21] R. Scherwitzl, S. Gariglio, M. Gabay, P. Zubko, M. Gibert, and J. M. Triscone, Phys. Rev. Lett. **106**, 246403 (2011).

[22] A. V. Boris *et al.*, Science **332**, 937 (2011).

[23] J. A. Liu, S. Okamoto, M. van Veenendaal, M. Kareev, B. Gray, P. Ryan, J. W. Freeland, and J. Chakhalian, Phys. Rev. B **83**, 161102 (2011).

[24] K. Yoshimatsu, T. Okabe, H. Kumigashira, S. Okamoto, S. Aizaki, A. Fujimori, and M. Oshima, Phys. Rev. Lett. **104**, 147601 (2010).

[25] J. Son, J. M. LeBeau, S. J. Allen, and S. Stemmer, Appl. Phys. Lett. **97**, 202109 (2010).

[26] L. F. Kourkoutis, J. H. Song, H. Y. Hwang, and D. A. Muller, Proc. Natl. Acad. Sci. **107**, 11682 (2010).

[27] J. M. LeBeau, S. D. Findlay, L. J. Allen, and S. Stemmer, Ultramicroscopy **110**, 118 (2010).

[28] J. M. LeBeau, A. J. D'Alfonso, N. J. Wright, L. J. Allen, and S. Stemmer, Appl. Phys. Lett. **98**, 052904 (2011).

[29] J. Hwang, J. Y. Zhang, J. Son, and S. Stemmer, Appl. Phys. Lett. **100**, 191909 (2012).

[30] E. J. Kirkland, *Advanced Computing in Electron Microscopy* (Springer, New York, 2010).

[31] J. L. García-Muñoz, J. Rodríguez-Carvajal, P. Lacorre, and J. B. Torrance, Phys. Rev. B **46**, 4414 (1992).

[32] P. Hohenberg, and W. Kohn, Phys. Rev. B **136**, B864 (1964).

[33] W. Kohn, and L. J. Sham, Phys. Rev. **140**, A1133 (1965).

[34] See supplemental material at [link to be inserted by publisher] for PACBED patterns as a function of octahedral tilt angles and a PACBED study of a $[LaNiO_3/SrTiO_3]_3$ superlattice.

[35] D. G. Ouellette, S. Lee, J. Son, S. Stemmer, L. Balents, A. J. Millis, and S. J. Allen, Phys. Rev. B **82**, 165112 (2010).

[36] M. K. Stewart *et al.*, J. Appl. Phys. **110**, 033514 (2011).

[37] J. S. Zhou, J. B. Goodenough, and B. Dabrowski, Phys. Rev. Lett. **94**, 226602 (2005).

[38] S. B. Lee, R. Chen, and L. Balents, Phys. Rev. B **84**, 165119 (2011).

[39] H. Park, A. J. Millis, and C. A. Marianetti, Phys. Rev. Lett. **109**, 156402 (2012).

**Table I.** Experimental (EXP) and DFT parameters for LaNiO$_3$ films, LaNiO$_3$ in superlattices, and bulk. All films and superlattices are coherently strained to LSAT. Experimental Ni-O-Ni bond angles and Ni-O bond distances in films and superlattices are calculated from the measured tilt angles.

| LaNiO$_3$ phase | Tilt angles $\alpha, \gamma$ (°) | Pseudocubic u.c. parameters: $a, c$ (Å) | Ni-O-Ni (°) Out-of-plane | Ni-O-Ni (°) In-plane | Ni-O (Å) Out-of-plane | Ni-O (Å) In-plane |
|---|---|---|---|---|---|---|
| Bulk (EXP)[a] | 5.2, 5.2 | 3.84 | 165.3 | 165.3 | 1.94 | 1.94 |
| Bulk (DFT) | 4.62, 4.62 | 3.76 | 166.8 | 166.8 | 1.892 | 1.892 |
| Single-layer film (EXP)[b] | 6.2 ± 0.4, 0.9 ± 0.8 | 3.87 ± 0.02, 3.82 ± 0.02 | 162 ± 1.1 | 167 ± 1.0 | 1.94 ± 0.01 | 1.95 ± 0.01 |
| Single-layer film (DFT) | 6.30, 0.08 | 3.79, 3.74 | 163.0 | 167.6 | 1.891 | 1.906 |
| Superlattice (bottom layer) (EXP) | 2.8 ± 0.4, 7.1 ± 0.6 | 3.87 ± 0.01, 3.91 ± 0.01 | 172 ± 1.1 | 165 ± 1.4 | 1.96 ± 0.0 | 1.96 ± 0.01 |
| Superlattice (bottom layer) (DFT) | 0.10, 7.98 | 3.79, 3.83 | 170.3 | 163.0 | 1.921 | 1.916 |
| Superlattice (top layer) (EXP) | 4.6 ± 0.8, 5.4 ± 1.4 | 3.87 ± 0.01, 3.87 ± 0.01 | 167 ± 2.3 | 166 ± 3.1 | 1.95 ± 0.01 | 1.95 ± 0.01 |

[a] Ref. [31]

[b] Ref. [29]

**Figure Captions**

**Figure 1:** Schematic showing 2×2 pseudocubic u.c.'s of LaNiO$_3$, and definition of the octahedral tilt angles in the Glazer notation.

**Figure 2:** (a) Cross-section HAADF-STEM image of a [LaNiO$_3$/SrTiO$_3$]$_2$ superlattice. (b) experimental PACBED patterns taken from each layer in the superlattice. The boxes areas in (a) indicate the area from which the PACBED pattern was obtained in each case. (c) Simulated PACBED patterns that resulted in the best-match with the experiment using the $\chi^2$ comparisons shown in (d). The TEM sample thickness yielding the best-match is indicated. (d) $\chi^2$ maps as a function of $\alpha$ and $\gamma$ tilt angles for each layer. The brightest pixel has the lowest $\chi^2$, and the contour lines are fits to the map. (e) Magnified portion of the PACBED pattern from the bottom LaNiO$_3$ showing the elongation along the *c*-axis (growth direction) of the pseudocubic u.c.

**Figure 3.** Schematic showing the mechanisms by which octahedral connectivity and matching of the lattice parameter to the substrate result in the observed tilt patterns and negative Poisson's ratio.

**Figure 4.** Inverse of the sheet resistance (1/$R_s$) as a function of superlattice repeats (*n*) for [LaNiO$_3$/SrTiO$_3$]$_n$ at different temperatures.

**Figure 1**

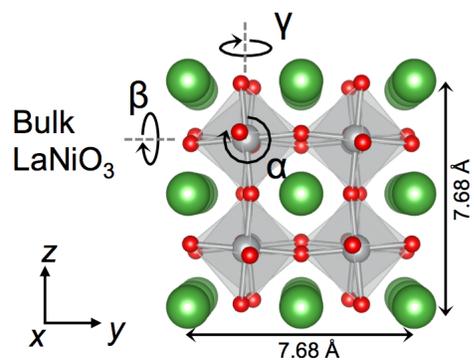

**Figure 2**

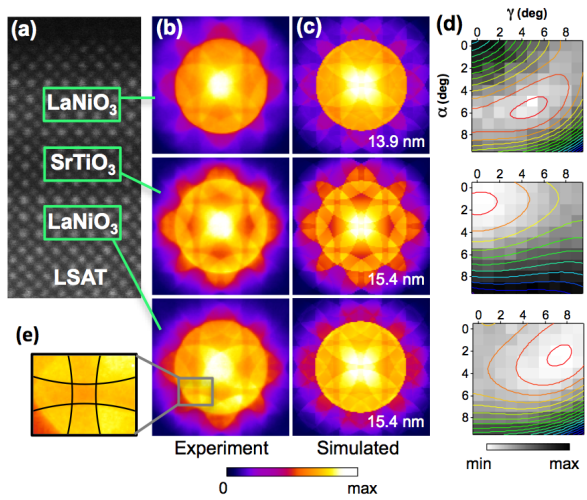

**Figure 3**

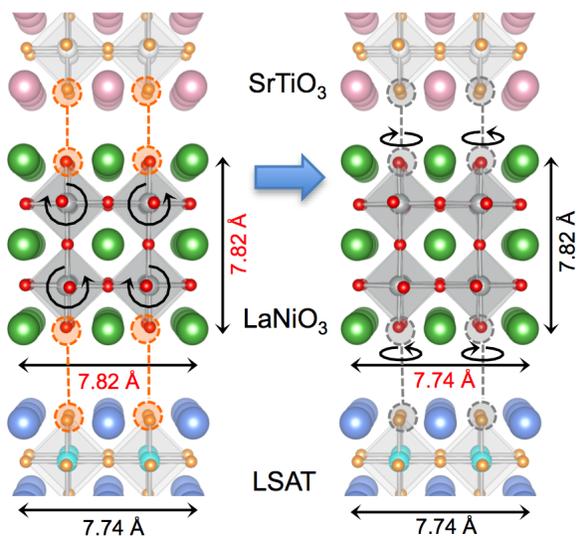

**Figure 4**

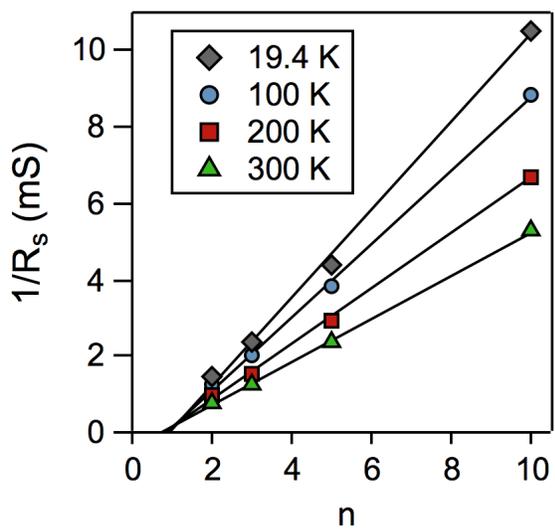